\documentclass{article}
\usepackage[latin1]{inputenc}
\usepackage[T1]{fontenc}
\usepackage{natbib}
\usepackage{times}
\usepackage{graphicx}
\usepackage{url}
\usepackage{multirow}
\usepackage{amsmath,amssymb,amsfonts}
\usepackage[margin=1in]{geometry} 

\newcommand {\br}[1]{\left(#1\right)}
\def\RR{\mathbb{R}}
\newcommand {\OMIT}[1]{}
\newcommand {\nm}[1]{\Arrowvert\, #1 \,\Arrowvert}
\newcommand {\inpH}[2]{\left\langle #1 \right\rangle_{#2}}

\begin{document}

\title{Spectral analysis of gene expression profiles using gene networks}
\author{Franck Rapaport\\Center for Computational Biology\\Ecole des Mines de Paris\\ and Service de Bioinformatique\\Institut Curie\\ \texttt{Franck.Rapaport@curie.fr} \and Andrei Zinovyev\\Service de Bioinformatique\\Institut Curie\\ \texttt{Andrei.Zinovyev@curie.fr} \and Marie Dutreix\\CNRS-UMR 2027\\Institut Curie\\ \texttt{Marie.Dutreix@curie.fr} \and Emmanuel Barillot\\Service de Bioinformatique\\Institut Curie\\ \texttt{Emmanuel.Barillot@curie.fr} \and Jean-Philippe Vert\\Center for Computational Biology\\Ecole des Mines de Paris\\ \texttt{Jean-Philippe.Vert@ensmp.fr}}

\maketitle

\begin{abstract}
Microarrays have become extremely useful for analysing genetic phenomena, but establishing a relation between microarray
analysis results (typically a list of genes) and their biological significance is often difficult. Currently, the standard approach
is to map \emph{a posteriori} the results onto gene networks to elucidate the functions perturbed at the level of pathways. However, integrating \emph{a priori} knowledge of
the gene networks could help in the statistical analysis of gene expression data and in their biological interpretation. Here we propose a method to integrate \emph{a priori} the knowledge of a gene network in the analysis of gene expression data. The approach is based on the spectral decomposition of gene expression profiles with respect to the eigenfunctions of the graph, resulting in an attenuation of the high-frequency components of the expression profiles with respect to the topology of the graph. We show how to derive unsupervised and supervised classification algorithms of expression profiles, resulting in classifiers with biological relevance. We applied the method to the analysis of a set of expression profiles from irradiated and non-irradiated yeast strains. It performed at least as well as the usual classification but provides much
more biologically relevant results and allows a direct biological interpretation.

Supplementary material: \texttt{http://bioinfo.curie.fr/projects/kernelchip}
\end{abstract}

\section{Introduction}

During the last decade microarrays have become the technology of
choice for dissecting the genes responsible for a phenotype. By
monitoring the activity of virtually all the genes from a sample in a
single experiment they offer a unique perspective for explaining the
global genetic picture of a variant, whether a diseased
individual or a sample subject to whatever stressing conditions.

However, this strength is also the major weakness of
microarrays, and has led to ''gene list'' syndrome. With careful
experimental design and data analysis, microarrays often
present a list of genes that are differentially expressed between
two conditions, or allow samples to be classified
according to their known phenotypic features. Once obtained, the meaning behind this list
of genes, being typically a few hundreds, must be deciphered. Normally, the biologist
relies on experience for making sense
out of the results. Microarrays are prone to certain
errors, and a single gene showing differential expression may not necessarily be involved in the process under
study. More fundamentally, when it is clear that a gene is involved, the biologist is often more
interested in the biological function than in the single gene. The behaviour of a biological function usually involves more than just the expression of a single
gene, but involves the entire set of genes that form the regulation
pathways governing the function, which have to be examined. 

One of the aims of microarray data analysis is to discover co-operating genes that are similarly affected during the experiment.
Many databases and tools help verify this \emph{a posteriori}. For example, Gene Ontology \citep{Consortium2000Gene}, Biocarta (www.biocarta.com), GenMAPP (www.genmapp.org) and KEGG \citep{Kanehisa2004KEGG}
all allow a list of genes to be crossed with genetic
networks, including metabolic, signalling or other regulation pathways. Basic
statistical analysis (see, e. g., \citealp{Hosack2003}) can then determine whether a pathway is
over-represented in the list, and whether it is over-activated or under-activated.
However, it is obvious that introducing information on the pathway
at this point in the analysis process sacrifices some statistical
power to the simplicity of the approach. For example, if we assume that genes in the same pathway show a coherent
expression pattern, assessing differential gene 
expression could be made more efficient; coherent expression patterns 
in a regulation pathway would be defined as those in which
over-expressed inhibitors or activators result in respectively under-expressed or overexpressed targets, or vice versa.
Thus, a small but coherent difference in the expression of all
the genes in a pathway should be more significant than
larger but random differences.

There is therefore a pressing need for
methods integrating \emph{a priori} pathway knowledge in the gene expression analysis process, and several attempts
have been carried out in that direction so far.
An initial approach is to derive a model for gene expression from an \emph{a priori} model of gene networks; for example, by deriving constraints on co-regulated gene expression. Logical discrete formalism
\citep{ThomasDLF2001} can be used to analyse all the possible
steady states of a biochemical reaction network described by positive and negative influences and can determine whether the observed gene expression may be explained by a perturbation of the network. If only the signs of the concentration differences between two steady states are considered, it is possible to solve the corresponding
Laplace equation in sign algebra \citep{Radulescu2005JBSI},
giving qualitative predictions for the signs of the concentration
differences measured by microarrays.  Other approaches, such as the MetaReg formalism
\citep{GatViks2004}, and boolean and Bayesian networks \citep{Akutsu2000Inferring,Friedman2000Using}  have also been used to predict possible gene expression patterns from
the network structure, although these approaches adhere less to the formal
theory of biochemical reaction networks. \OMIT{The principal source of
complexity in these models is associated with presence of feedback
loops.}

Unfortunately, methods based on network models are rarely satisfactory because detailed quantitative knowledge of the complete reaction network parameters is often lacking, 
or there are only fragments of the network structure available.
In these cases,
more phenomenological approaches need to be used. Pathway
scoring methods try to detect perturbated "modules" or
network pathways while ignoring the detailed network
topology (for recent reviews see \citealp{Curtis2005,Cavalieri2005}). It is assumed that the genes
inside a module are co-ordinately expressed, and thus a
perturbation is likely to affect many of them.
\OMIT{
Another category of methods compares expression of a set of genes
with a model expression pattern: potentially, a good model should
improve sensitivity of the analysis. In the context of large scale
cancer studies of gene expression, in \citep{Segal2004},
\citep{Segal2005} global gene expression profile was decomposed
into modules where every module exhibits a specific pattern of
expression in different cancer types and presumably has common set
of regulators. A notion of {\it metagene} as a set of genes
together with weights (positive and negative) assigned to every
gene was introduced to describe various modes of expression
variation in a gene set (see \citep{Bild2003}, \citep{Bild2006}).
The metagenes can be empirically estimated from data by using PCA-
and ICA-like techniques \citep{Brunet2004}.
}

With available databases containing tens of thousands reactions and
interactions (KEGG \citep{Kanehisa2004KEGG}, TransPath
\citep{TransPath2006}, BioCyc \citep{BioCyc2005}, Reactome
\citep{Reactome2005} and others), the problem is how to integrate the detailed graph of gene interactions (and not just crude characteristics such as the inter/intra-module
connectivity) into the core microarray data analysis. Some promising results
have been reported with regard to this problem. \citet{Vert2003Bioinformatics} developed a method for correlating interaction
graphs and different types of quantitative data, and \citet{Rahnenfuhrer2004} showed that explicitly taking the pathway distance
between pairs of genes into account enhances the statistical scores when identifying activated pathways. \citet{Hanisch2002Co-clustering} reported the co-clustering of gene expression and gene networks, and the PATIKA project
\citep{Babur2004} proposed quantifying the compatibility of a
pathway with a given microarray data by scoring individual
interactions.

In this paper, we investigate a different approach for integrating genetic networks early in the gene expression analysis. We propose a method for calculating the eigen modes of
the response of the gene network to a perturbation and suggest how these can be introduced
into supervised and unsupervised microarray data analysis. The
method uses a hierarchical approach to automatically divide the network into modules having
co-ordinated responses. We are able to filter out high-frequency noisy modes, thus increasing
our ability to interpret expression profiles from the 
underlying genetic network. We illustrate the relevance of our approach by analysing a gene expression dataset that monitors the transcriptional response of irradiated and non-irradiated yeast colonies \citep{Mercier2004Biological}. We show that by filtering out 80\% of the eigen modes of the KEGG metabolic network in the gene expression, we obtain accurate and interpretable discriminative model that may lead to new biological insights.

\section{Methods}
In this section, we will explain how a gene expression can be decomposed into the eigen modes of a gene network, and how to derive unsupervised and supervised classification algorithms from this decomposition.

\subsection{Spectral decomposition of gene expression profiles}
We consider a finite set of genes $V$ of cardinality $|V|=n$. The available gene network is represented by an undirected graph $G=(V,E)$ without loop and multiple edges, in which the set of vertices $V$ is the set of genes and $E\subset V\times V$ is the list of edges. We will use the notation $u\sim v$ to indicate that two genes $u$ and $v$ are neighbors in the graph, that is, $(u,v) \in E$. For any gene $u$, we denote the degree of $u$ in the graph by $d_u$, that is, its neighbour number. Gene expression profiling gives a value of expression $f(u)$ for each gene $u$, and is therefore represented by a function $f:V\rightarrow \RR$.

The Laplacian of the graph $G$ is the $n\times n$ matrix \citep{Chung1997Spectral}:
\begin{equation}\label{eq:laplacian}
\forall u,v \in V, \quad L(u,v) =
\begin{cases}
d_u &\text{if }u=v\,,\\
-1 &\text{if }u\sim v\,,\\
0 &\text{otherwise\,.}
\end{cases}
\end{equation}
The Laplacian is a central concept in spectral graph theory \citep{Mohar1997Some} and shares many properties with the Laplace operator on Riemannian manifolds. $L$ is known to be symmetric positive definite and singular. We denote its eigenvalues by $0=\lambda_1\leq \ldots \leq \lambda_n$ and the corresponding eigenvectors by $e_1,\ldots,e_n$. The multiplicity of $0$ as an eigenvalue is equal to the number of connected components of the graph, and the corresponding eigenvectors are constant on each connected component.  The eigen-basis of $L$ forms a Fourier basis and a natural theory for Fourier analysis and spectral decomposition on graphs can thus be derived \citep{Chung1997Spectral}. Essentially, the eigenvectors with increasing eigenvalues tend to vary more abruptly on the graph, as the smoothest functions (constant on each connected component) are associated with the smallest (zero) eigenvalue. In particular, the Fourier transform $\hat{f} \in \RR^n$ of any expression profile $f$ is defined by:
\[
\hat{f}_i = \sum_{u\in V} e_i(u)f(u),\quad i=1,\ldots,n\,.
\]

Like its continuous counterpart, the discrete Fourier transform can be used for smoothing or for extracting features. Here, our hypothesis is that analysing a gene expression profile from its Fourier transform on an \emph{a priori} given gene network should allow the profile to be further analysed through this prior knowledge. In the next two sections we illustrate the potential applications of this approach by describing how this leads to a natural definition for distances between expression profiles, and how this distance can be used for classification or regression purposes.

\subsection{Deriving a metric for expression profiles}
The definition of new metrics on expression profiles that incorporate information encoded in the graph structure is a first possible application of the spectral decomposition. Following the classical methodology in Fourier analysis, we assume that the signal captured in the low-frequency component of the expression profiles contains the most biologically relevant information, particularly the general expression trends, whereas the high-frequency components are more likely measurement noise. For example, the low frequency component of an expression vector on the gene metabolic network should reveal areas of positive and negative expression on the graph that are likely to correspond to the activation or inhibition of pathways. We can translate this idea mathematically by considering the following class of transformations for expression profiles:
\begin{equation}\label{eq:transformation}
\forall f \in \RR^V,\quad S_\phi(f) = \sum_{i=1}^n \hat{f}_i \phi(\lambda_i) e_i \,,
\end{equation}
where $\phi:\RR^+ \rightarrow \RR$ is a non-increasing function that quantifies how each frequency is attenuated. For example, if we take $\phi(\lambda) = 1$ for all $\lambda$, the profile does not change that is, $S_\phi(f)=f$. However, if we take:
\begin{equation}\label{eq:phithres}
\phi_{\text{thres}}(\lambda) = \begin{cases}
1 & \text{if }0\leq\lambda\leq\lambda_0\,,\\
0 & \text{if }\lambda > \lambda_0\,,
\end{cases}
\end{equation}
we produce a low-pass filter that removes all frequencies from $f$ above the threshold $\lambda_0$. Finally, a function of the form:
\begin{equation}\label{eq:phiexp}
\phi_{\exp}(\lambda) = \exp(-\beta \lambda)\,,
\end{equation}
 for some $\beta >0$, keeps all the frequencies but strongly attenuates the high-frequency components.

If $S_\phi(f)$ includes the biologically relevant part of the expression profile, we can compare two expression profiles $f$ and $g$ through their representations $S_\phi(f)$ and $S_\phi(g)$. This leads to the following metric between the profiles:
\[
\begin{split}
d_\phi(f,g)^2 &= \nm{S_\phi(f) - S_\phi(g)}^2 \\
&=\sum_{i=1}^n \br{\hat{f}_i- \hat{g}_i}^2 \phi(\lambda_i)^2\,.
\end{split}
\]
The associated norm is the Euclidean norm associated with the inner product:
\[
\begin{split}
\inpH{f,g}{\phi} &= \sum_{i=1}^n \hat{f}_i \hat{g}_i \phi(\lambda_i)^2\\
&= \sum_{i=1}^n f^\top e_i e_i^\top g \phi(\lambda_i)^2\\
&= f^\top K_\phi g\,,
\end{split}
\]
where $K_\phi = \sum_{i=1}^n \phi(\lambda_i)^2 e_i e_i^\top$ is the positive semidefinite matrix obtained by modifying the eigenvalues of $L$ through $\phi$. For example, taking $\phi(\lambda)=\exp\br{-\beta\lambda}$ leads to $K_\phi = \exp_M\br{-\beta L}$, where $\exp_M$ denotes the matrix exponential.

\subsection{Supervised learning and regression}
The construction of predictive models for a property of interest from the gene expression profiles of the studied samples is a second possible application of the spectral decomposition of expression profiles on the gene network. Typical applications include predicting cancer diagnosis or prognosis from gene expression data, or discriminating between different treatments applied to micro-organisms. Most approaches build predictive models from the gene expression alone, and then check whether the predictive model is biologically relevant by studying, for example, whether genes with high weights are located in similar pathways. However the genes often give no clear biological meaning. Here, we propose a method combining both steps in a single predictive model that is trained by forcing some form of biological relevance.

We use linear predictive models to predict a variable of interest $y$ from an expression profile $f$ that are obtained by solving the following optimisation problem:
\begin{equation}\label{eq:minreg}
\underset{w \in \RR^n}{\min} \sum_{i=1}^p l(w^\top f_i , y_i) + C \nm{w}^2\,,
\end{equation}
where $(f_1,y_1),\ldots,(f_p,y_p)$ is a training set of profiles containing the variable $y$ to be predicted, and $l$ is a loss function that measures the cost of predicting $w^\top f_i$ instead of $y_i$. For example, the popular support vector machine \citep{Boser1992training} is a particular case of equation (\ref{eq:minreg}) in which $y$ can take values in ${-1,+1}$ and $l(u,y) = max(0,1-yu)$ is the hinge loss function; ridge regression is obtained for $y \in \RR$ by taking $l(u,y) = (u-y)^2$ \citep{Hastie2001elements}.

Here, we do not apply algorithms of the form (\ref{eq:minreg}) directly to the expression profiles $f$, but to their images $S_\phi(f)$. That is, we consider the problem:
\begin{equation}\label{eq:minregphi}
\underset{w \in \RR^n}{\min} \sum_{i=1}^p l(w^\top S_\phi(f_i) , y_i) + C \nm{w}^2\,.
\end{equation}
Indeed, for any $w \in \RR^d$, let $v=K_\phi^{1/2} w$. We observe that for any $f \in \RR^n$:
\[
\begin{split}
w^\top S_\phi(f) &= w^\top \sum_{i=1}^n \hat{f}_i \phi(\lambda_i) e_i \\
&= f^\top \sum_{i=1}^n e_i \phi(\lambda_i) e_i^\top w \\
&= f^\top K_\phi^{1/2} w\\
&= f^\top v\,,
\end{split}
\]
showing that the final predictor obtained by minimizing (\ref{eq:minregphi}) is equal to $v^\top f$. We also note that:
\[
\begin{split}
\nm{w}^2 &= w^\top w\\
&= v^\top K_\phi^{-1} v\\
&= \sum_{i=1}^n \frac{\hat{v}_i^2}{\phi(\lambda_i)^2}\,,
\end{split}
\]
where the last equality remains valid if $K_\phi$ is not invertible simply by not including in the sum the terms $i$ for which $\phi(\lambda_i)=0$. This shows that (\ref{eq:minregphi}) is the equivalent of solving the following problem in the original space:
\begin{equation}\label{eq:minregphiorig}
\underset{v \in \RR^n}{\min} \sum_{i=1}^p L(v^\top f_i , y_i) + C  \sum_{i:\phi(\lambda_i)>0} \frac{\hat{v}_i^2}{\phi(\lambda_i)^2}\,.
\end{equation}
Thus, the resulting algorithm amounts to finding a linear predictor $v$ that minimizes the loss function of interest $l$ regularised by a terms that penalises the high-frequency components of $v$. This is different to the classical regularisation $\nm{v}^2$ that only focuses on the norm of $v$. As a result, the linear predictor $v$ can be made smoother on the gene network by increasing the parameter $C$. This allows the prior knowledge to be direcly included because genes in similar pathways would be expected to contribute similarly to the predictive model.

There are two consequences of this procedure. First, if the true predictor really is smooth on the graph, the formulation (\ref{eq:minregphiorig}) can help the algorithm focus on plausible models even with very little training data, resulting in a better estimation. As a result, we can expect a better predictive performance. Second, by forcing the predictive model $v$ to be smooth on the graph, biological interpretation of the model should be easier by inspecting the areas of the graph in which the predictor is strongly positive or negative. Thus the model should be easier to interpret than models resulting from the direct optimisation of equation (\ref{eq:minreg}).

\section{Data}

We collected the expression data from a study analysing the effect  of low irradiation doses on \emph{Saccharomyces cerevisiae} strains \citep{Mercier2004Biological}. The first group of extracted expression profiles was a set of twelve independent yeast cultures grown without radiation (not irradiated, NI). From this group, we excluded an outlier that was indicated to us by the author of the article. The second group was a set of six independent irradiated (I) cultures exposed to a dose of 15-20 mGu/h for 20h. This dose produces no mutagenic effects, but induces transcriptional changes. We used the same normalisation method as in the first study of this data (Splus LOWESS function, see \citealp{Mercier2004Biological} for details) and attempted (1) to separate the NI samples from the I ones, and (2) to understand the difference between the two populations in terms of metabolic pathways.

The gene network model used to analyse the gene expression data was therefore built from the KEGG database of metabolic pathways \citep{Kanehisa2004KEGG}. The metabolic gene network is a graph in which the enzymes are vertices and the edges between two enzymes indicate that the product of a reaction catalysed  by the first enzyme is the substrate of the reaction catalysed by the second enzyme. We reconstructed this network from the KGML v0.3 version of KEGG, resulting in 4694 edges between 737 genes. We kept only the largest connected component (containing 713 genes) for further spectral analysis.

\section{Results}

\subsection{Unsupervised classification}

\begin{figure*}
\centerline{\includegraphics[width=9cm]{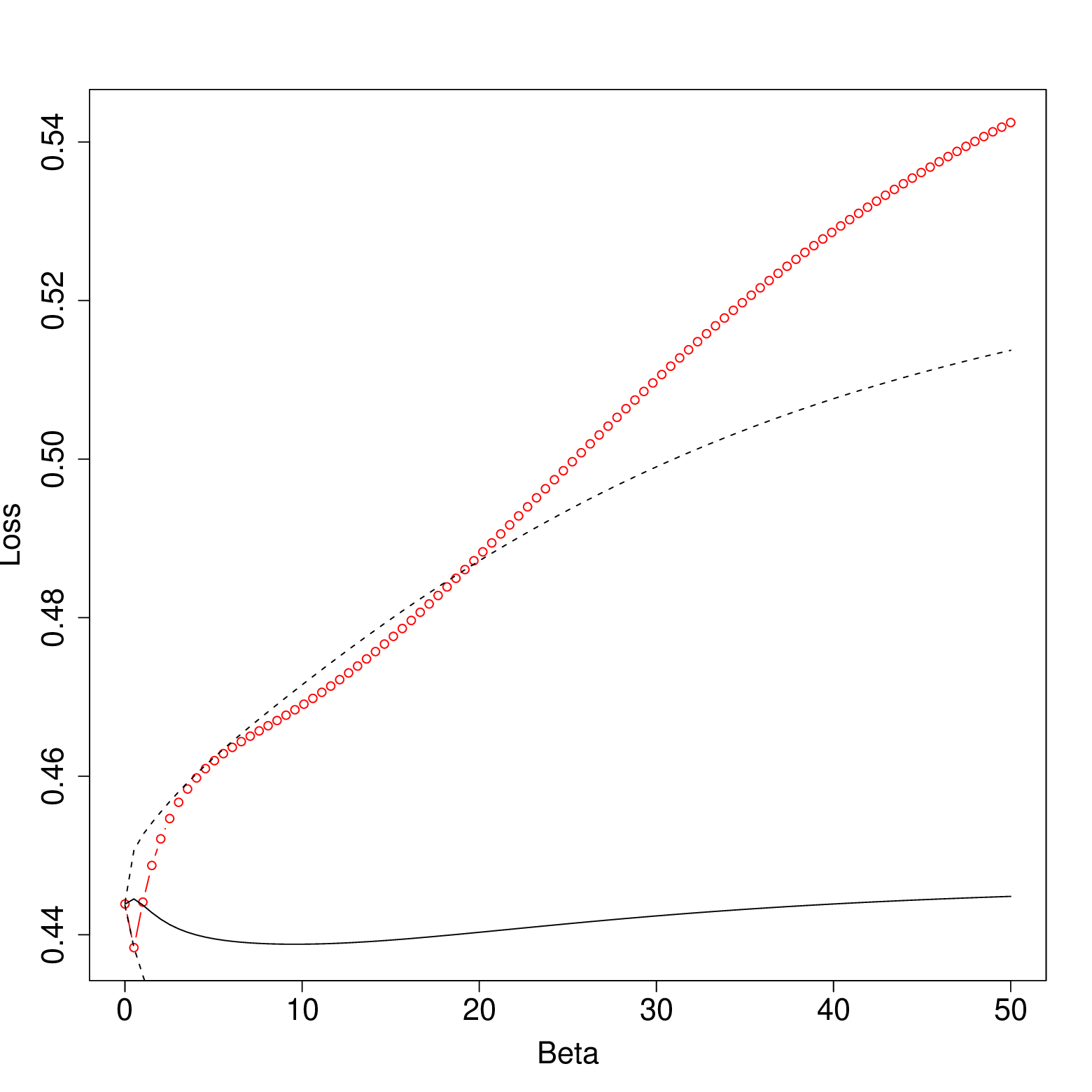}\includegraphics[width=9cm]{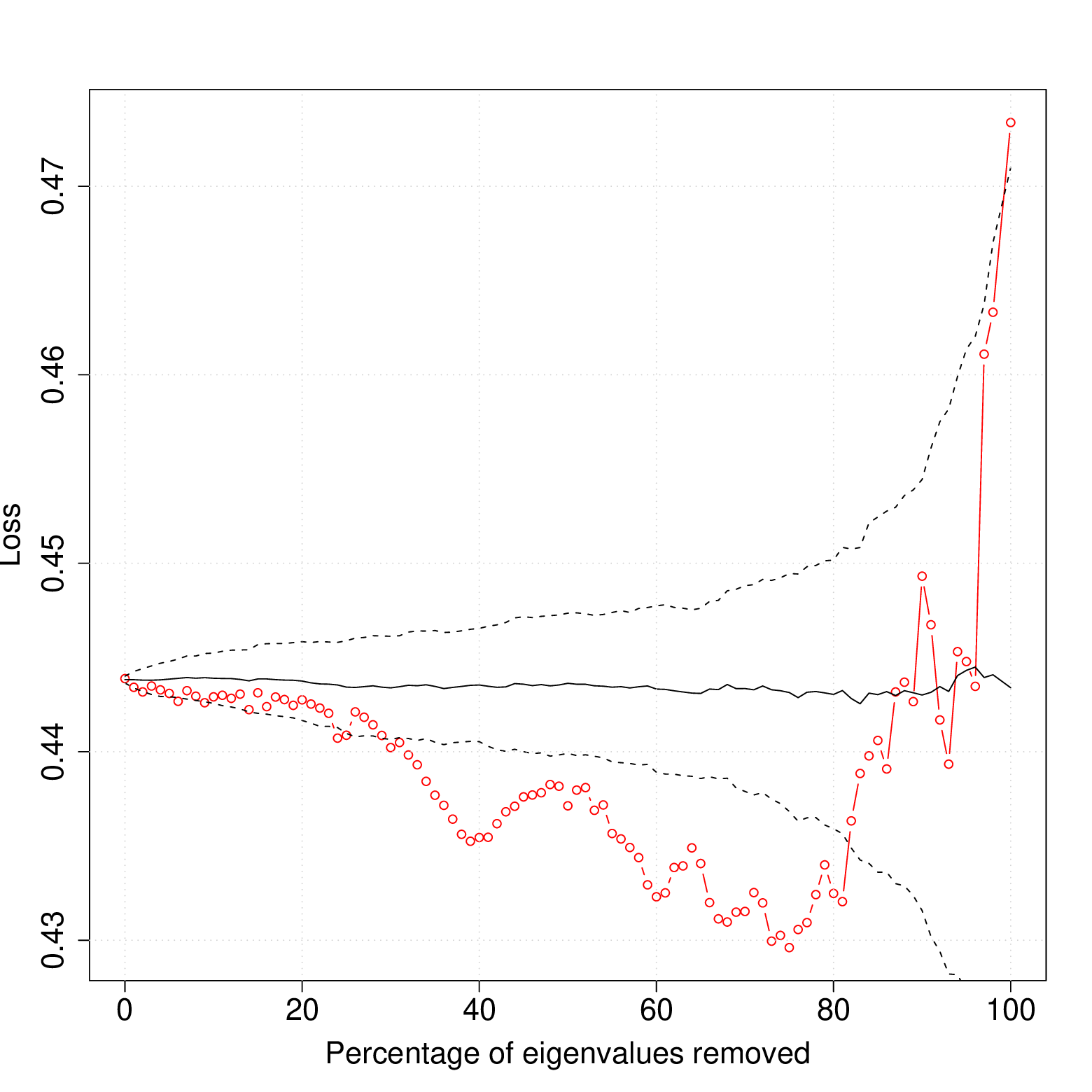}}
\caption{Performance of the unsupervised classification after changing the metric with the function $\phi(\lambda)=\exp(-\beta\lambda)$ for different values of $\beta$ (left), or with the function $\phi(\lambda)=1(\lambda<\lambda_0)$ which varying $\lambda_0$, that is, by keeping a variable number of smallest eigenvalues (right).. The red curve is obtained with the KEGG network. The black curves show the result (mean and one standard deviation interval) obtained with a random network.}\label{fig:unsupervised_beta}
\end{figure*}

First, we tested the general effect of modifying the distances between expression profiles using the KEGG metabolic pathways as background information in an unsupervised setting. We calculated the pairwise distances between all $17$ expression profiles after applying the transformations defined by the filters  (\ref{eq:phithres}) and (\ref{eq:phiexp}), over a wide range of parameters. We assessed whether the resulting distances were more coherent with a biological intepretation by calculating the ratio of intraclass distances over all pairwise distances, defined by:
\[
r = \frac{\sum_{u_1,v_1 \in V_1} d(u_1,v_1)^2 + \sum_{u_2,v_2 \in V_2}{d(u_2,v_2)^2}}{\sum_{u,v \in V}d(u,v)^2}\,,
\]
where $V_{1}$ and $V_{2}$ are the two classes of points. We compared the results with those obtained by replacing KEGG with a random network, produced by keeping the same graph structure but randomly permutating the vertices, in order to assess the significance of the results. We generated 100 such networks to give an average result with a standard deviation. Figure \ref{fig:unsupervised_beta} shows the result for the funtion $\phi_{\exp}(\lambda)=\exp(-\beta\lambda)$ with varying $\beta$ (left), and for the function $\phi_{\text{thres}}(\lambda)=1(\lambda<\lambda_0)$ for varying $\lambda_{0}$ (right). We found that, apart from very small values of $\beta$, the change of metric with the $\phi_{\exp}$ function performed worse than that of a random network. The second method (filtering out the high frequency components of the gene expression vector), in which up to 80\% of the eigenvectors were removed, performed significantly better than that of a random network. When only the top $3\%$ of the smoothest eigenvectors were kept, the performance was similar to that of a random network, and when only the top $1\%$ was kept, the performance was significantly worse. This explains the disappointing results obtained with the $\phi_{\exp}$ function: by giving more weight to the small eigenvalues exponentially, the method focuses on those first few eigenvectors that, as shown by the second method, do not provide a geometry compatible with the separation of samples into two classes. From the second plot, we can infer that at least 20\% of the KEGG eigenvectors should be given sufficient weight to obtain a geometry compatible with the classification of the data in this case.

\subsection{PCA analysis}

We carried out a Principal Component Analysis (PCA, \citealp{Jolliffe1996Principal}) on the original expression vectors $f$ and compared this with a PCA of the
transformed set of vectors $S_\phi(f)$ obtained with the function $\phi_{\text{thres}}$ to further investigate the effect on their relative positions of filtering out the high frequencies of the expression profiles.

\begin{figure*}
\centerline{a)\includegraphics[width=8cm]{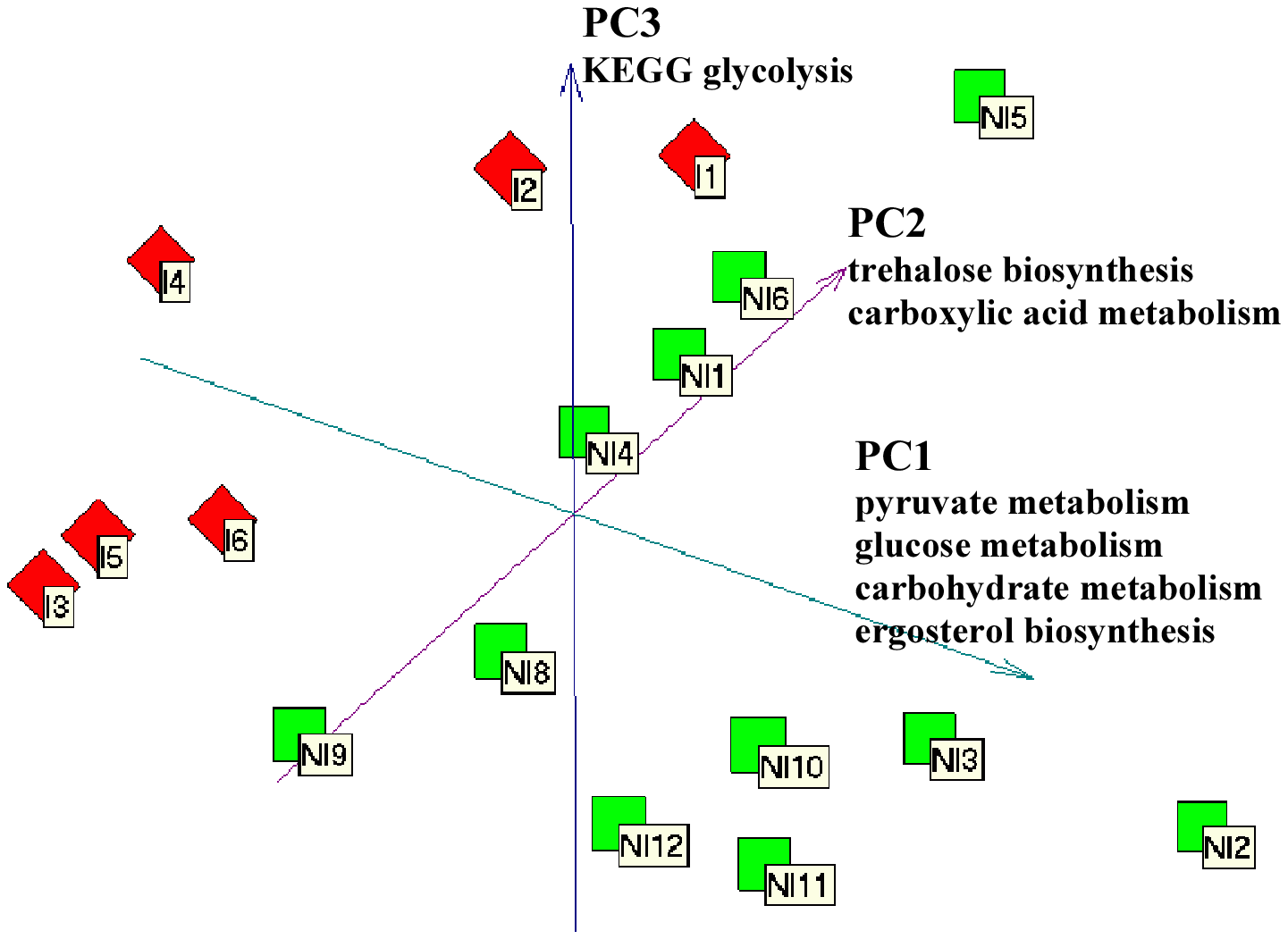}
b)\includegraphics[width=8.5cm]{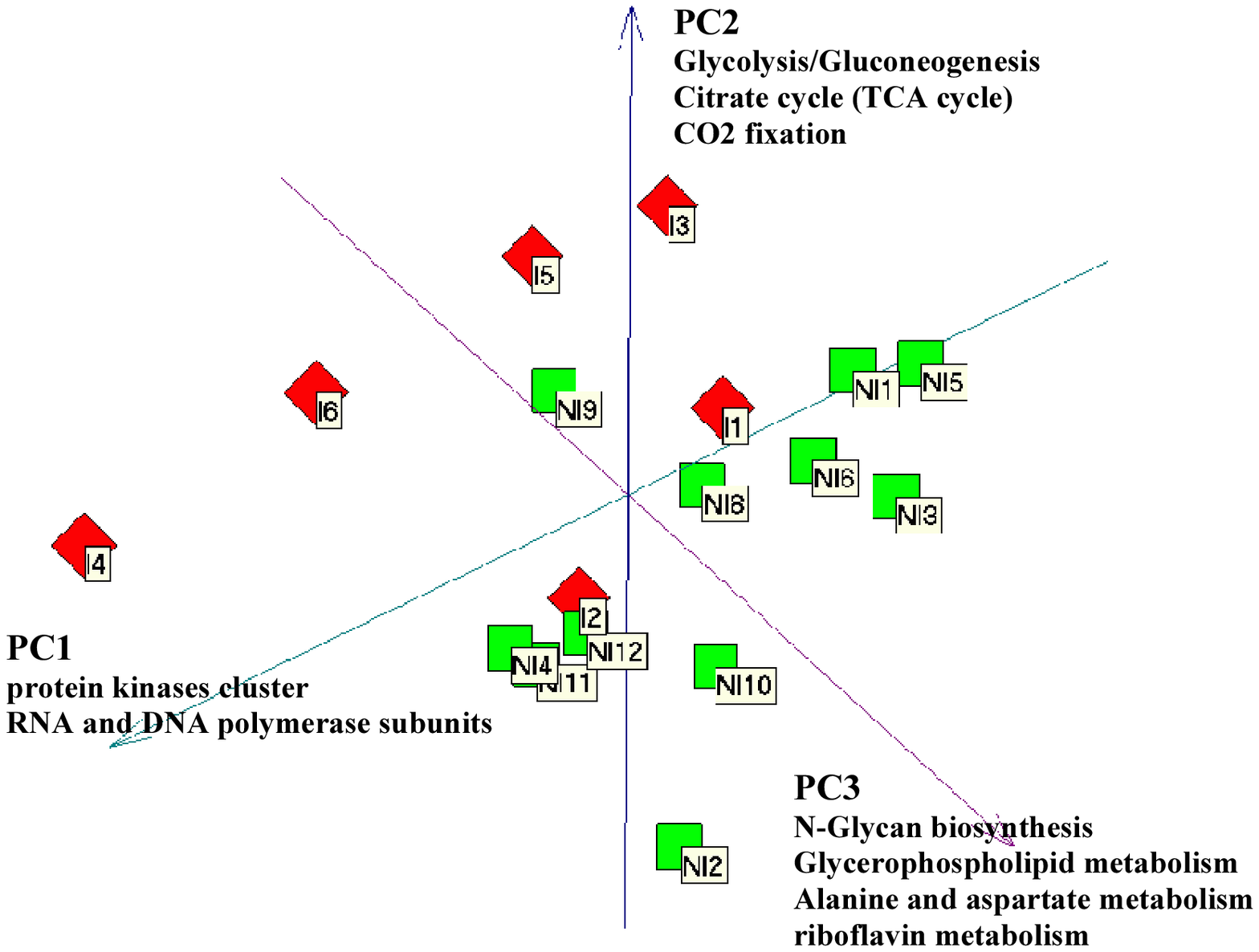}}
\caption{PCA plots of the initial expression profiles (a) and
the transformed profiles using network topology (80\% of the
eigenvalues removed) (b). The green squares are non-irradiated
samples and the red rhombuses are irradiated samples. Individual sample
labels are shown together with GO and KEGG annotations associated
with each principal component.}\label{fig:pcasample}
\end{figure*}

Analysis of the initial sample distribution (see
fig.\ref{fig:pcasample}) showed that the first principal component
can partially separate irradiated from non-irradiated samples,
with exception of the two irradiated samples "I1" and "I2". They
have larger projections onto the third principal component than
onto the first one. \OMIT{(JP: too early in the text) Further analysis showed that their profiles
are significantly different from all other irradiated samples so
that they were systematically misclassified.} The experimental
protocol revealed that these two samples were affected by higher
doses of radiation than the four other samples.

Gene Ontology analysis of the genes that gave the biggest
contribution in the first principal component showed that
"pyruvate metabolism", "glucose metabolism", "carbohydrate
metabolism", and "ergosterol biosynthesis" ontologies (here we
list only independent ontologies) were over-represented (with
p-values less than $10^{-10}$). The second component was
associated with "trehalose biosynthesis", and "carboxylic acid
metabolism" ontologies and the third principal component was
associated with the KEGG glycolysis pathway. The first three
principal components collected 25\%, 17\% and 11\% of the total
dispersion.

The transformation (\ref{eq:transformation}) resulting from a step-like
attenuation of eigenvalues $\phi_{\text{thres}}$ removing 80\% of the largest
eigenvalues significantly changed the global layout of data
(fig. \ref{fig:pcasample}, right) but generally preserved the local
neighbourhood relationships. The first three principal components
collected 28\%, 20\% and 12\% of the total dispersion, which was
only slightly higher than the PCA plot of the initial profiles.
The general tendency was that the non-irradiated normal samples
were more closely grouped, which explains the lower intraclass
distance values shown in fig.\ref{fig:unsupervised_beta}. The
principal components in this case allowed them to be associated
with gene ontologies with higher confidence (for the first
component, the p-values are less than $10^{-25}$). This is a direct consequence of the fact that the principal components are constrained to belong to a subspace of smooth functions on KEGG, giving coherence in terms of pathways to the genes contributing to the components. The first
component gave "DNA-directed RNA polymerase activity", "RNA
polymerase complex" and "protein kinase activity".
Fig.\ref{fig:cytoscape} shows that these are the most connected
clusters of the whole KEGG network. The second component is
associated with "purine ribonucleotide metabolism", "RNA
polymerase complex", "carboxylic acid metabolism" and "acetyl-CoA
metabolism" ontologies and also with "Glycoly\-sis/Glu\-coneogenesis",
"Citrate cycle (TCA cycle)" and "Reductive carboxylate cycle (CO2
fixation)" KEGG pathways. The third component was associated with
"prenyltransferase activity", "lyase activity" and "aspartate
family amino acid metabolism" ontologies and with "N-Glycan
biosynthesis", "Glycerophospholipid metabolism", "Alanine and
aspartate metabolism" and "riboflavin metabolism" KEGG pathways.
Thus, the PCA components of the transformed expression profiles
were affected both by network features and by the microarray data.

\subsection{Supervised classification}

We tested the performance of supervised classification after
modifying the distances with a support vector machine (SVM) trained
to discriminate irradiated samples from non-irradiated samples. For each
change of metric, we estimated the performance of the SVM from the total
number of misclassifications and the total hinge loss using a ''leave-one-out'' (LOO) approach. This approach removes each sample in turn, trains a classifier on the remaining samples and then tests the resulting classifier on the removed sample. For each fold, the regularization parameter was selected from the training
set only by minimising the classification error estimated with an internal
LOO experiment. The calculations were carried out using the \texttt{svmpath} package in the \texttt{R} computing
environment.

\begin{figure*}
\centerline{\includegraphics[width=9cm]{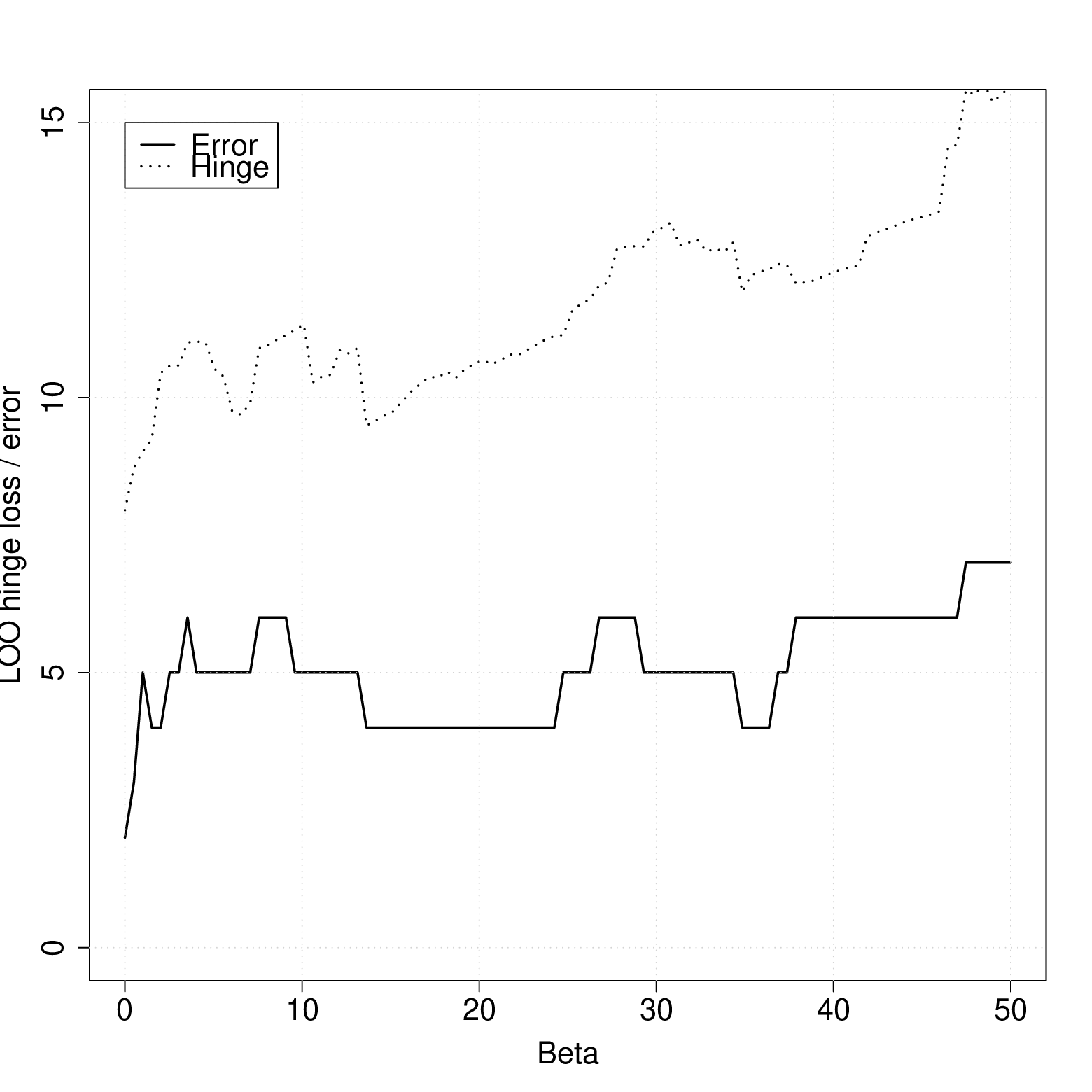}\includegraphics[width=9cm]{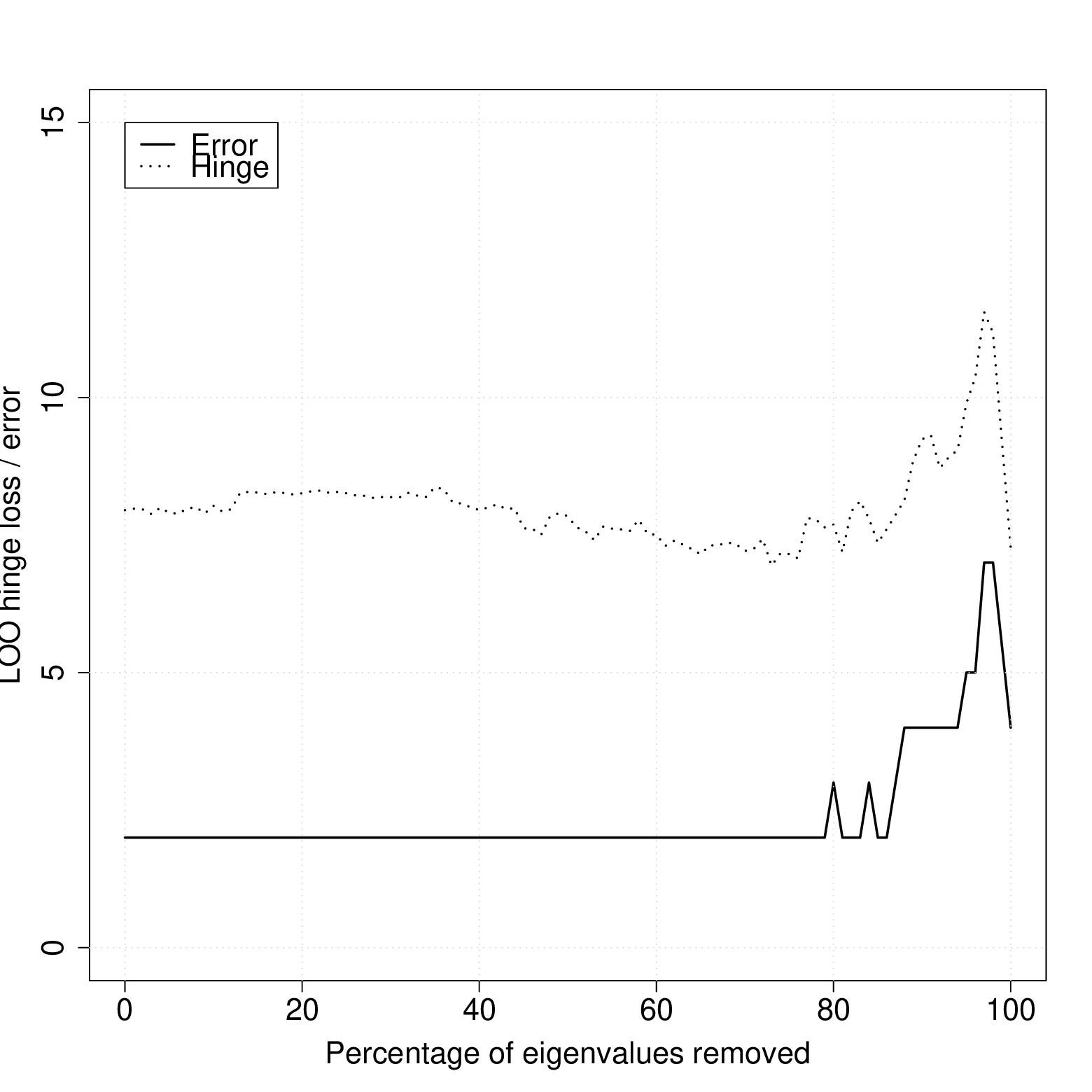}}
\caption{Performance of the supervised classification when changing the metric with the function $\phi_{\exp}(\lambda)=\exp(-\beta\lambda)$ for different values of $\beta$ (left picture), or the function $\phi_{\text{thres}}(\lambda)=1(\lambda < \lambda_{0})$ for different values of $\lambda_0$ (i.e., keeping only a fraction of the smallest eigenvalues, right picture). The performance is estimated from the number of misclassifications in a leave-one-out error.}\label{fig:supervised}
\end{figure*}

Figure \ref{fig:supervised} shows the classification results for the
two high frequency attenuation functions $\phi_{\exp}$ and $\phi_{\text{thres}}$ with varying
parameters. The baseline LOO error was $2$ misclassifications for
the SVM in the original Euclidean space. For the exponential
variant ($\phi_{\exp}(\lambda)=\exp(-\beta\lambda)$), we observed an irregular
but certain degradation in performance for positive $\beta$ for both
the hinge loss and the misclassification number. This is consistent with the result shown in fig.\ref{fig:unsupervised_beta} in which the
change of metric towards the first few eigenvectors does not give a geometry coherent with the
classification of samples into irradiated and non-irradiated,
resulting in a poorer performance in supervised classification as well.
For the second variant, in which the expression profiles were projected onto the eigenvector of the
graph with the smallest eigenvalues, we found that the
performance remained as accurate as the baseline performance until up to
$80\%$ of the eigenvectors were discarded, with the hinge loss
even exhibiting a slight minimum in this region. This is consistent with the classes being more clustered in this case
than in the original Euclidean space. Overall these results show
 that classification accuracy can be kept high even when the classifier is constrained to exhibit a certain coherence with
the graph structure.

\subsection{Interpretation of the SVM classifier}
\begin{figure*}
\centerline{\includegraphics[width=10cm]{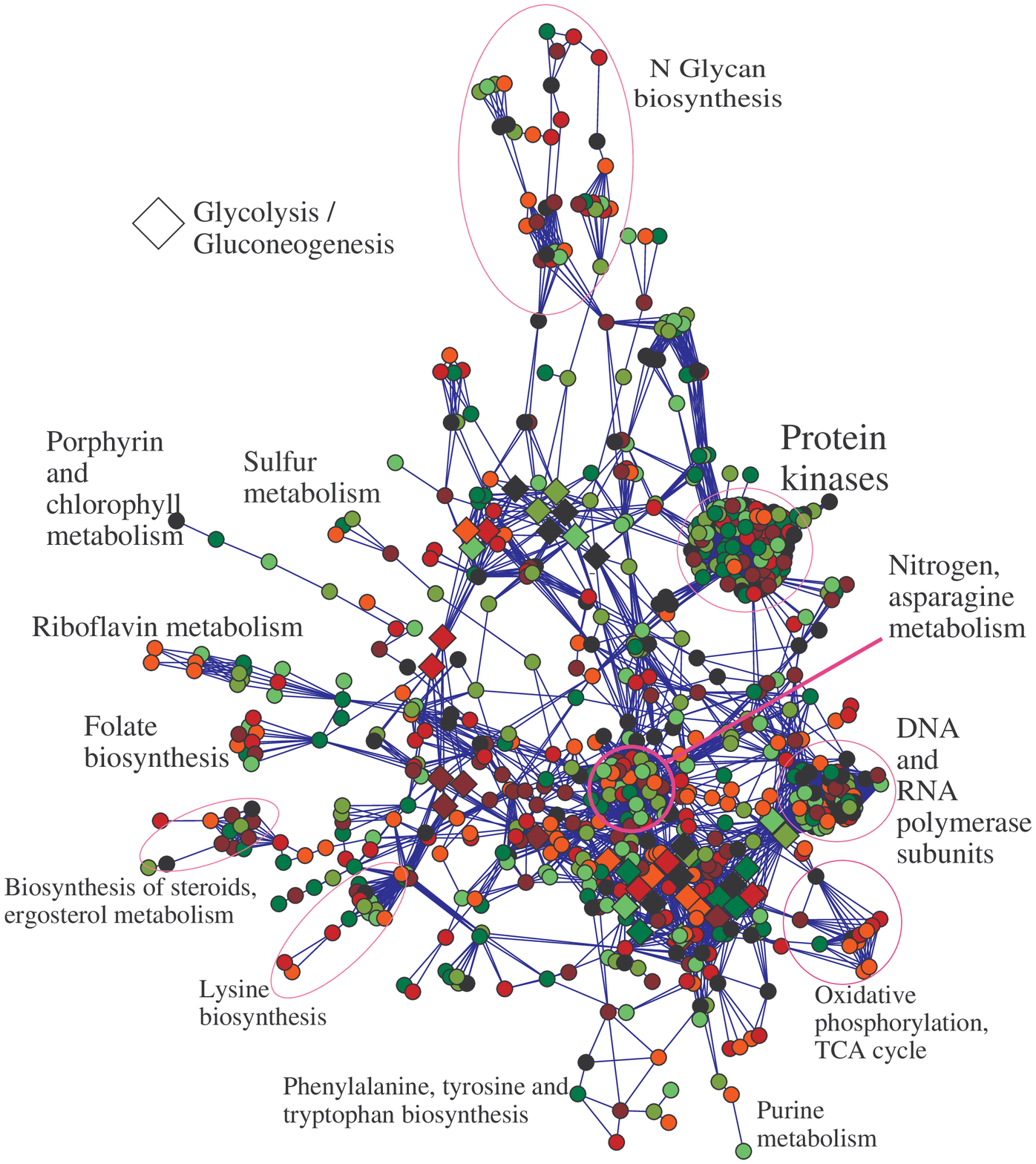}
\includegraphics[width=7.3cm]{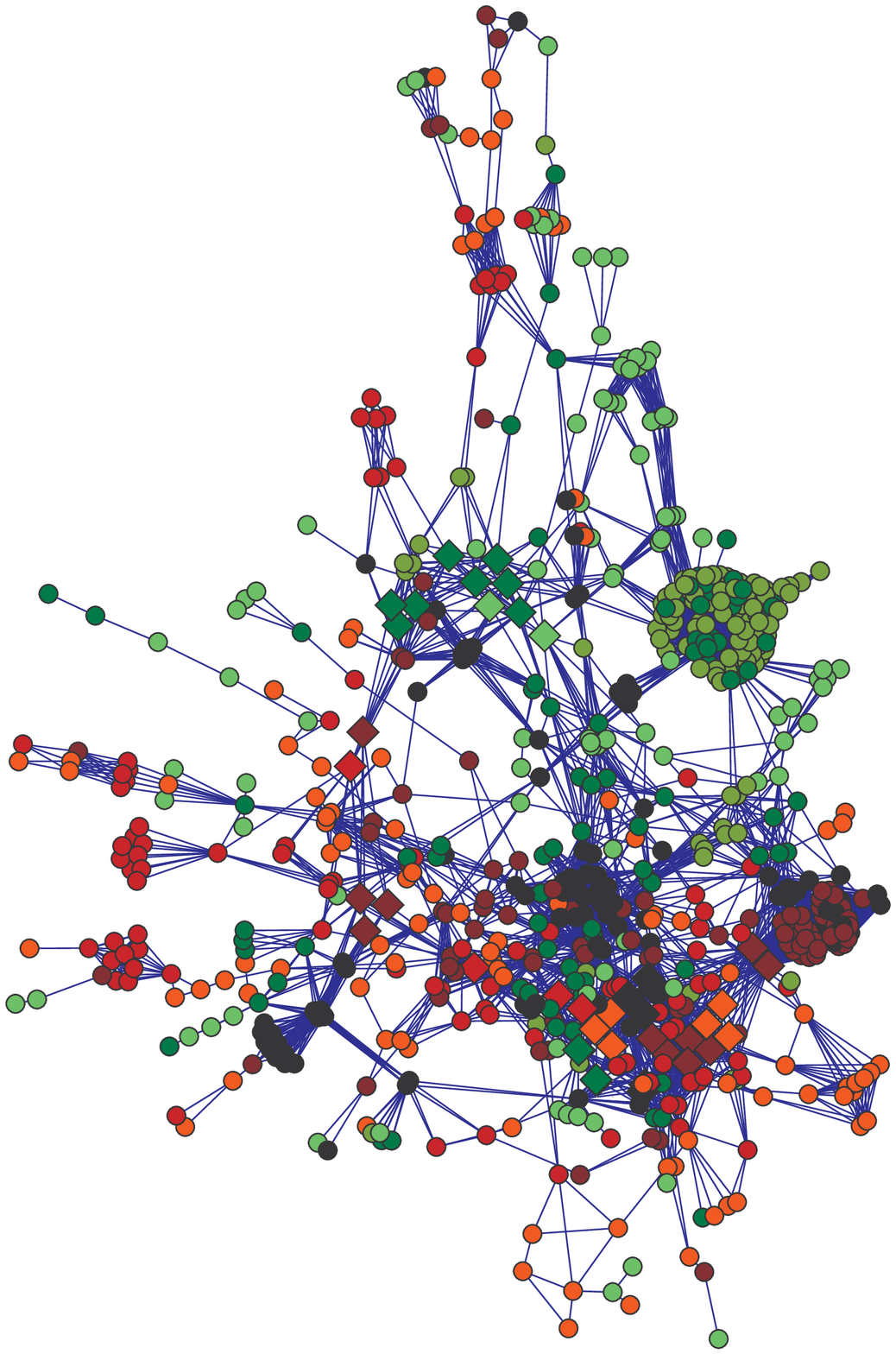}}
\caption{Global connection map of KEGG with mapped coefficients of
the decision function obtained by applying a customary linear SVM
\textbf{(left)} and using high-frequency eigenvalue attenuation
(80\% of high-frequency eigenvalues have been removed)
\textbf{(right)}. Spectral filtering divided the whole network
into modules having coordinated responses, with the activation of
low-frequency eigen modes being determined by microarray data.
Positive coefficients are marked in red, negative
coefficients are in green, and the intensity of the colour
reflects the absolute values of the coefficients. Rhombuses
highlight proteins participating in the Glycolysis/Gluconeogenesis
KEGG pathway. Some other parts of the network are annotated
including big highly connected clusters corresponding to protein
kinases and DNA and RNA polymerase sub-units. }\label{fig:cytoscape}
\end{figure*}

Figure (\ref{fig:cytoscape}) shows the global connection map of
KEGG generated from the connection matrix by Cytoscape
software \citep{Cytoscape2003}. The coefficients of the decision
function $v$ of equation (\ref{eq:minregphiorig}) for the
classifier constructed either in the original Euclidean space or after filtering the $80\%$ top spectral components of the expression profiles are shown in colour. We used a color scale from green (negative weights) to red (positive weights) to provide an easy visualization of the main features of the classifiers. Both classifiers give the same classification error but the classifier
constructed using the network structure can be more naturally
interpreted, because the classifier variables are grouped according
to their participation in the network modules.

Although from a biological point of view, very little can be learned
from the classifier obtained in the original Euclidian space (figure \ref{fig:cytoscape}, left), it
is indeed possible to distinguish several features of interest for the classifier obtained in the second case (figure \ref{fig:cytoscape},
right). First, oxidative phosphorylation
was found among 
the pathways with the most positive weights, which is consistent with previous analyses showing that this pathway tended to be up-regulated after irradiation \citep{Mercier2004Biological}. An important cluster
involving the DNA and RNA polymerases is also found to bear weights slightly above average in these experiments. Several studies have previously reported the induction of genes involved in replication and repair after high doses of irradiation \citep{Mercier2001Transcriptional}, but the detection of such
an induction at the low irradiation doses used in the
present biological experiments is rather interesting.
The strongly negative landscape of weights in the protein kinases cluster has not been seen before and may lead to a new area of biological study. Most of the kinases are involved in signalling pathways, and therefore their low expression levels may have important biological consequences.

\begin{figure*}
\centerline{\includegraphics[width=8cm]{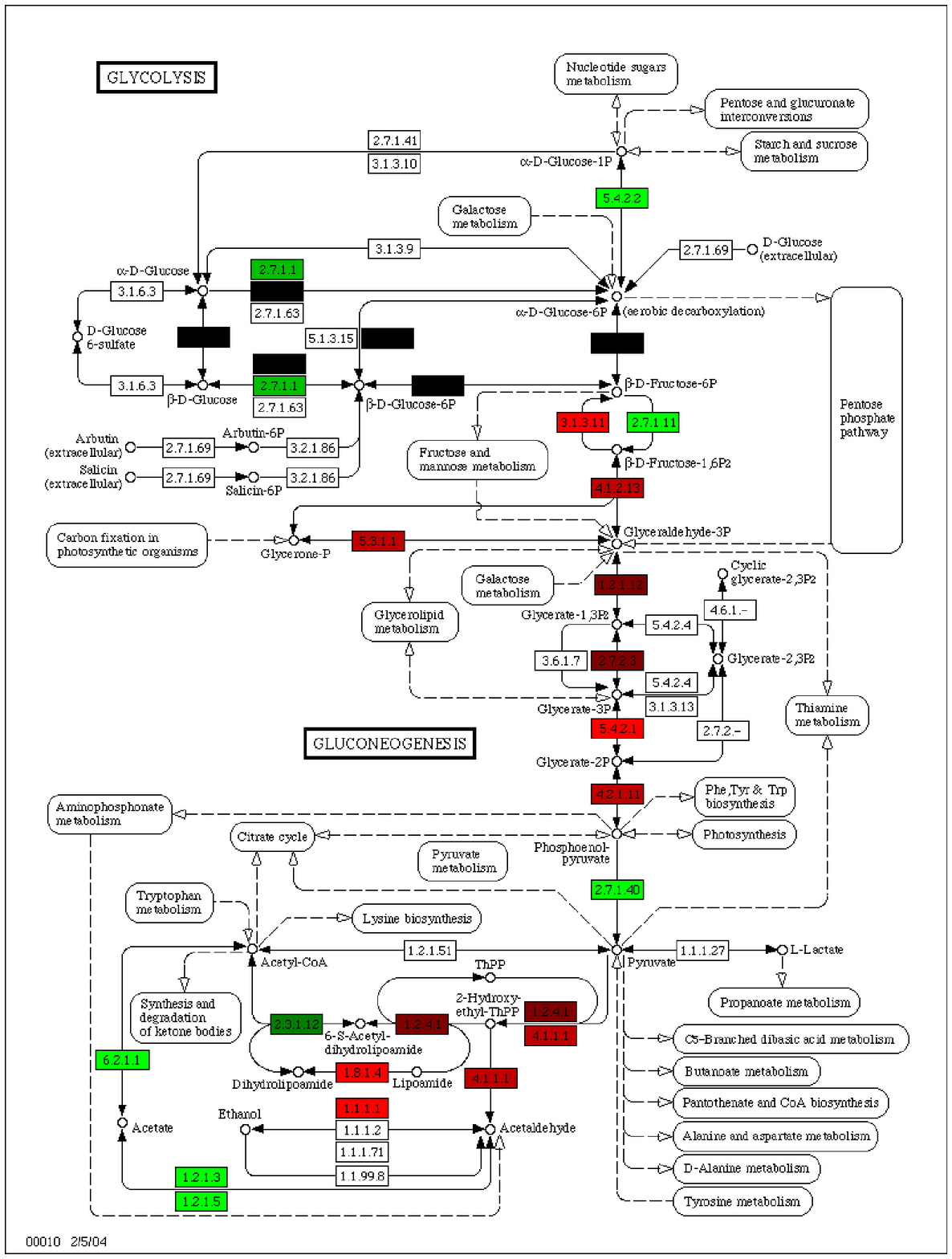}a)\hspace{1cm}
     \includegraphics[width=8cm]{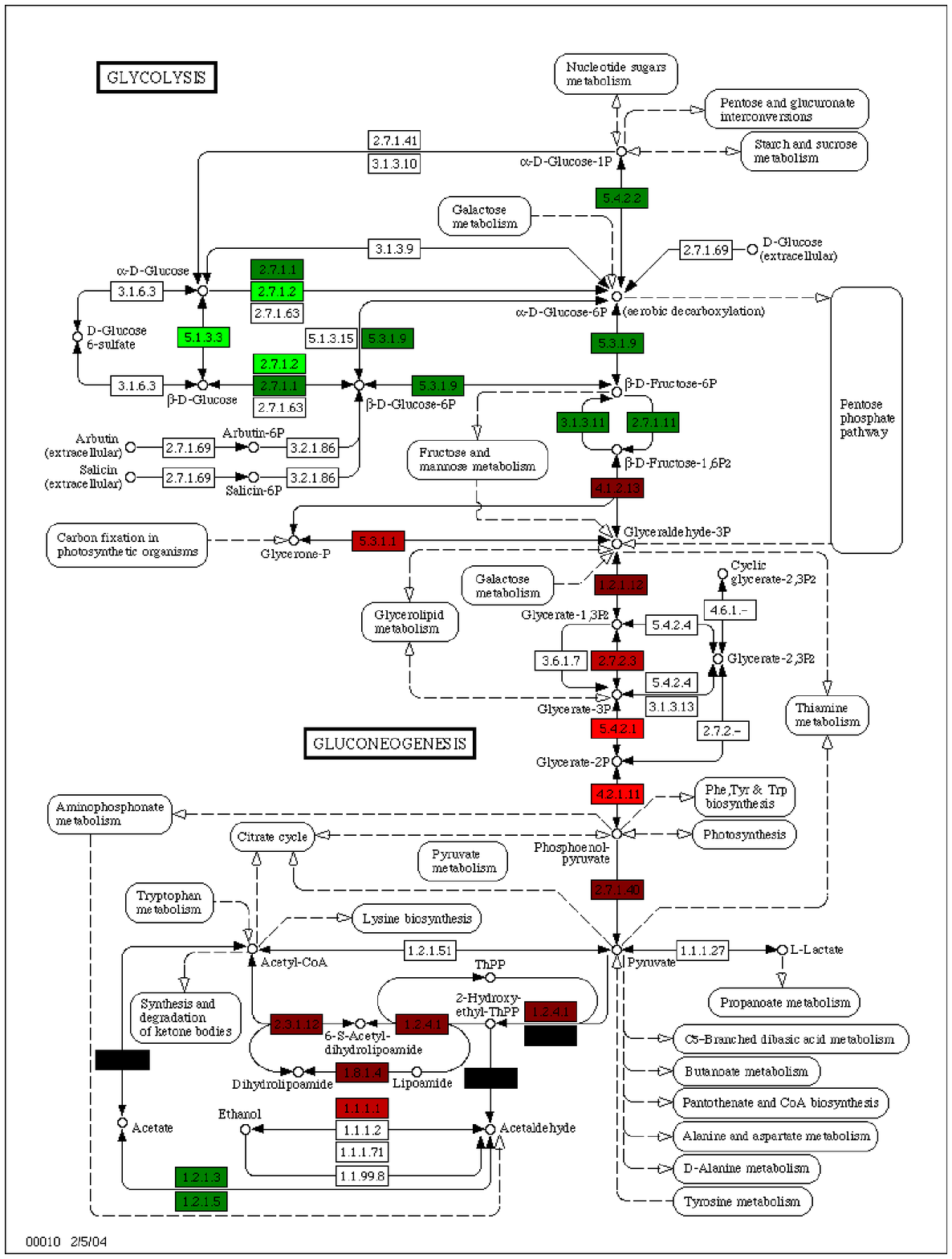}b)}
\caption{The glycolysis/gluconeogenesis pathways of KEGG with
mapped coefficients of the decision function obtained by applying
a customary linear SVM  \textbf{(a)} and using high-frequency
eigenvalue attenuation \textbf{(b)}. The pathways are mutually
exclusive in a cell, as clearly highlighted by our algorithm.}
\label{fig:glucpathways}
\end{figure*}

Figure \ref{fig:cytoscape} shows a highlighted
pathway named "Glycoly\-sis/Gluconeogenesis" in KEGG. A more detailed view of this pathway is shown in fig.\ref{fig:glucpathways}. This pathway contains enzymes that are also used in many other KEGG pathways and is therefore situated in the middle and most entangled part of the global
network. As already mentioned, this pathway is associated
with the first and the third principal components of the initial
dataset. The pathway actually contains two alternative sub-pathways
that are affected differentially. Over-expression in the gluceogenesis pathway seems to be characteristic of irradiated samples, whereas glyceolyses has a low level of expression in that case .This shift can be
observed by changing from anaerobic to aerobic growth conditions
(called diauxic shift). The reconstruction of this
from our data with no prior input of this knowledge strongly
confirms the relevance of our analysis method. It also shows that analysing expression in terms of the global up- or down-regulation of entire pathways as defined, for example, by KEGG, could mislead as there are many antagonist processes that take place inside pathways. Representing KEGG as a large network helps keep the biochemical relationships between genes without the constraints of pathway limits.

\section{Discussion}

Our algorithm groups predictor variables according to
highly connected "modules" of the global gene
network. We assume that the genes within a tightly connected network module are likely to contribute similarly to the prediction
function because of the interactions between the genes. This motivates the filtering of
gene expression profile to remove the noisy high-frequency modes
of the network.

Such grouping of variables is a very useful feature of the
resulting classification function because the function becomes
meaningful for interpreting and suggesting biological
factors that cause the class separation. This allows
classifications based on functions, pathways and network
modules rather than on individual genes. This can lead to a more robust behaviour of the classifier in
independent tests and to equal if not better classification results.
Our results on the dataset we analysed shows only a slight improvement, although this may be due to its limited size.
Therefore we are currently extending our work to larger data sets.

An important remark to bear in mind when analyzing pictures such as fig.\ref{fig:cytoscape} and \ref{fig:glucpathways} is that the colors represent the weights of the classifier, and not gene expression levels. There is of course a relationship between the classifier weights and the typical expression levels of genes in irradiated and non-irradiated samples: irradiated samples tend to have expression profiles positively correlated with the classifier, while non-irradiated samples tend to be negatively correlated. Roughly speaking, the classifier tries to find a smooth function that has this property. If more samples were available, better non-smooth classifier might be learned by the algorithm, but constraining the smoothness of the classifier is a way to reduce the complexity of the learning problem when a limited number of samples are available. This means in particular that the pictures provide virtually no information regarding the over- or under-expression of individual genes, which is the cost to pay to obtain instead an interpretation in terms of more global pathways. Constraining the classifier to rely on just a few genes would have a similar effect of reducing the complexity of the problem, but would lead to a more difficult interpretation in terms of pathways.

An advantage of our approach over other
pathway-based clustering methods is that we consider the network modules that naturally appear from spectral analysis
rather than a historically defined separation of the network into
pathways. Thus, pathways cross-talking is taken into account, which is difficult to do using other approaches. It can however be noticed that the implicit decomposition into pathways that we obtain is biased by the very incomplete knowledge of the network and that certain regions of the network are better understood, leading to a higher connection concentration.

Like most approaches aiming at comparing expression data with gene networks such as KEGG, the scope of this work is limited by two important constraints. First the gene network we use is only a convenient but rough approximation to describe complex biochemical processes; second, the transcriptional analysis of a sample can not give any information regarding post-transcriptional regulation and modifications. Nevertheless, we believe that our basic assumptions remain
valid, in that we assume that the expression of the genes belonging
to the same metabolic pathways module are coordinately regulated. Our interpretation of the results supports this
assumption.

Another important caveat is that we simplify the network description as
an undirected graph of interactions. Although this would seem to be
relevant for simplifying the description of metabolic networks, real gene regulation networks are influenced by the
direction, sign and importance of the interaction. Although the incorporation of weights into the Laplacian  (equation \ref{eq:laplacian}) is straightforward and allows the extension of the approach to weighted undirected graphs, the incorporation of directions and signs to represent signalling or regulatory pathways requires more work but could lead to important advances for the interpretation of microarray data in cancer studies, for example.

\section*{Acknowledgments}

This work was supported by the grant ACI-IMPBIO-2004-47 of the French Ministry for Research and New Technologies. We thank Sabrina Carpentier from the Service de Bioinformatique of the Institut Curie for the help she provided with the normalisation of the microarray data.

\bibliographystyle{plainnat}


\end{document}